
\magnification=\magstep1
\vbadness=10000
\parskip=\baselineskip
\parindent=10pt
\centerline{\bf MATTERS OF GRAVITY}
\bigskip
\bigskip
\line{Number 4 \hfill Fall 1994}
\bigskip
\bigskip
\bigskip
\centerline{\bf Table of Contents}
\bigskip
\hbox to 6truein{Editorial {\dotfill} 2}
\hbox to 6truein{Correspondents {\dotfill} 2}
\bigskip
\hbox to 6truein{\bf Gravity news:\hfill}
\hbox to 6truein{Report on the APS Topical Group in Gravitation,
Beverly Berger   {\dotfill} 3}
\bigskip
\hbox to 6truein{\bf Research briefs:\hfill}
\hbox to 6truein{Gravitational Microlensing and the Search for
Dark Matter:\hfil}
\hbox to 6truein{A Personal View, Bohdan Paczynski  {\dotfill} 5}
\smallskip
\hbox to 6truein{Laboratory Gravity:  The G Mystery  \hfill}
\hbox to 6truein{and Intrinsic Spin Experiments,
Riley Newman {\dotfill} 11}
\smallskip
\hbox to 6truein{LIGO Project update, Stan Whitcomb {\dotfill} 15}
\bigskip
\hbox to 6truein{\bf Conference Reports:\hfill}
\smallskip
\hbox to 6truein{PASCOS '94, Peter Saulson {\dotfill} 16}
\smallskip
\hbox to 6truein{The Vienna meeting, P. Aichelburg, R. Beig {\dotfill} 19}
\smallskip
\hbox to 6truein{The Pitt Binary Black Hole Grand Challenge Meeting,
Jeff Winicour {\dotfill} 20}
\smallskip
\hbox to 6truein{International Symposium on Experimental Gravitation,
\hfil}
\smallskip
\hbox to 6truein{Nathiagali, Pakistan, Munawar Karim {\dotfill} 22}
\smallskip
\hbox to 6truein{10th Pacific Coast Gravity Meeting
, Jim Isenberg
{\dotfill} 23}
\bigskip
\bigskip
\bigskip
\bigskip
\leftline{\bf Editor:}

\medskip
\leftline{Jorge Pullin}
\smallskip
\leftline{Center for Gravitational Physics and Geometry}
\leftline{The Pennsylvania State University}
\leftline{University Park, PA 16802-6300}
\smallskip
\leftline{Fax: (814)863-9608}
\leftline{Phone (814)863-9597}
\leftline{Internet: pullin@phys.psu.edu}

\vfill
\eject

\centerline{\bf Editorial}

The newsletter strides on. I had to perform some pushing around and
arm-twisting to get articles for this number. I wish to remind everyone
that suggestions and ideas for contributions are {\bf especially}
welcome. The newsletter is growing rather weak on the theoretical
side. Keep those suggestions coming!

I put together this newsletter mostly on a palmtop computer while
travelling, with some contributions arriving the very day of
publication (on which, to complicate matters, I was giving a talk at a
conference, my email reader crashed and our network went down).  One
of the contributions is a bit longer than the usual format.  Again it
is my fault for failing to warn the author in due time. I apologize
for this and other errors.

The next newsletter is  due February 1st.  Many thanks to the authors and
the correspondents who made this issue possible.

If everything goes well this newsletter should be available in the
gr-qc Los Alamos bulletin board under number gr-qc/9409004. To
retrieve it send email to gr-qc@xxx.lanl.gov (or gr-qc@babbage.sissa.it
in Europe) with Subject: get 9409004 (issue 2 is available as
9309003 and issue 3 as 9402002). All issues are available as postscript
files in the WWW http://vishnu.nirvana.phys.psu.edu/  Or email me. Have fun.

\medskip
\leftline{Jorge Pullin}

\bigskip
\bigskip
\centerline{\bf Correspondents}
\medskip

\parskip=2pt
\item{1.} John Friedman and Kip Thorne: Relativistic Astrophysics,
\item{2.} Jim Hartle: Quantum Cosmology and Related Topics
\item{3.} Gary Horowitz: Interface with Mathematical High Energy Physics,
    including String Theory
\item{4.} Richard Isaacson: News from NSF
\item{5.} Richard Matzner: Numerical Relativity
\item{6.} Abhay Ashtekar and Ted Newman: Mathematical Relativity
\item{7.} Bernie Schutz: News From Europe
\item{8.} Lee Smolin: Quantum Gravity
\item{9.} Cliff Will: Confrontation of Theory with Experiment
\item{10.} Peter Bender: Space Experiments
\item{11.} Riley Newman: Laboratory Experiments
\item{12.} Peter Michelson: Resonant Mass Gravitational Wave Detectors
\item{13.} Stan Whitcomb: LIGO Project
\parskip=\baselineskip

\vfill
\eject

\centerline{\bf Report on the APS Topical Group in Gravitation}
\medskip
\centerline{Beverly Berger, Oakland University}
\centerline{berger@vela.oakland.edu}
\bigskip
\bigskip

\noindent As most of you already know, a petition to form a Topical Group
in Gravitation (TGG) in the Americal Physical Society (APS) has
been circulating since early February.  The objective of the TGG
is to provide a distinct focus within the APS for research in
gravitational physics.  The existence of major projects such as
LIGO makes the formation of such an entity especially important
now.  At this moment, we have 165 of the 200 signatures required
to form the TGG.  The final effort to reach the goal is now
underway.  I urge any of you who are APS members and have not
yet signed the petition to please do so now. (A copy accompanies
this note.)
\medskip

\noindent The next stage in the process will be the formation of an {\it
ad hoc} organizing committee (AHOC) to draw up bylaws and to
serve as a nominating committee for officers of the TGG (as
required by APS).  Several of you have already volunteered for
the AHOC.  I welcome any others who are interested.
\medskip

\noindent Please stay tuned for future developments.

\eject

\centerline{PETITION TO THE COUNCIL OF THE AMERICAN PHYSICAL SOCIETY}
\medskip

\noindent {We, the undersigned members of the American Physical
Society,
petition the Council of the American Physical Society to establish a
Topical Group in Gravitation.  Areas of interest to the proposed Topical
Group include, but are not limited to, experiments and observations
related to the detection and interpretation of gravitational waves,
experimental tests of gravitational theories, computational general
relativity, relativistic astrophysics, solutions to Einstein's equations
and their properties, alternative theories of gravity, classical and
quantum cosmology, and quantum gravity.  The purpose of the Topical
Group is to provide a unified forum for these areas of current research
which now span several Divisions of the Society.}
\medskip
\medskip

\centerline{Signature\hskip 3cm Name (Printed) \hskip 3cm Affiliation}
\bigskip
\bigskip
\bigskip
\hrule
\bigskip
\bigskip
\hrule
\bigskip
\bigskip
\hrule
\bigskip
\bigskip
\hrule
\bigskip
\bigskip
\hrule
\bigskip
\bigskip
\hrule
\bigskip
\bigskip
\hrule
\bigskip
\bigskip
\hrule
\bigskip
\bigskip
\hrule

\bigskip

\medskip

\noindent Return to:

\centerline{Beverly K. Berger}

\centerline{Physics Department}

\centerline{Oakland University}

\centerline{Rochester, MI  48309}

\vfill
\eject

\centerline{\bf Gravitational Microlensing and the Search for Dark Matter}
\centerline{A Personal View}
\medskip
\centerline{Bohdan Paczy\'nski, Princeton University}
\centerline{bp@astro.princeton.edu}
\bigskip
\bigskip
\medskip

Gravitational microlensing is a phenomenon based on the General Theory
of Relativity: any mass concentration, a black hole, a star, a brown
dwarf or a planet, distorts space time and changes the direction
of the light rays, acting like a lens.  The prediction that stars
observed near the eclipsed solar disk should be displaced by almost
2 arc-seconds was verified observationally be the 1919 expedition
headed by Sir A. Eddington.

In the following decades the same theoretical discovery was made independently
every few years, by prominent physicists and astrophysicists as well
as by little known waiter from Brooklyn: if the observer and any
two stars in our own galaxy are well aligned then two images of the
more distant star will be formed by the gravitational field of the
closer star acting as a lens.  One can make an estimate that the
alignement should be better than one part in $ 10^8 $, i.e. the
angular separation between the two stars should be less than a milli
arc second.  And the problem is that there are no telescopes capable
of resolving such small angles.  The other problem: such a precise
alignement is very improbable.

Eight years ago I made pretty much the same discovery for the tenth
time (Paczy\'nski 1986).  The two independent referees pointed out
correctly that there was nothing original in my paper, it was just
a compilation of fragments of various ideas.  Somehow I managed to
persuade the Editor to accept the paper.  As it turned out we were
all wrong.  There was one very new and important ingredient in my paper,
the magic words: ``dark matter''.  Without realizing it I bridged the
inter-disciplinary gap.  A physicist from the nearby Jadwin Hall at
Princeton, Dave Bennet, became a frequent visitor to my office at
Peyton Hall.  He was very seriously interested in the problem of dark matter.

The problem existed for about half a century.  There was evidence
from the dynamics of galaxies and their clusters that there is plenty
of matter out there that cannot be accounted for with the stars that
shine (Hammond 1994).  Was the dark matter made of elementary particles
with a mass of $ 10^{-5} $ eV, or supermassive black holes of
$ 10^6 ~ M_{\odot} $ (one million solar masses) each, was anybody's
guess.  Dave Bennett was one of the many who were eager to solve the
mystery.  My paper proposed a specific way of conducting the search:
all it takes is to monitor the brightness of a few million stars
for a few years.  Whenever any object more massive than planet Earth
moves in front of one of those stars a double image forms.
Even though the two images would be seen as one, their total brightness
would change in a predictable way.  For the lens as massive
as the sun the characteristic time scale for the so called
microlensing event is a month or two.  And other things being
equal the time scale is proportional to the square root of mass of
the lensing object.  And millions of stars are needed because the
expected rate of events is very small.

The principle was fine, but it was considered a science fiction by
me as well as all other astronomers, with a possible exception of
Ken Freeman from Australia.  Dave Bennet was not an astronomer and he
did not know that there are all those variable stars that would form
a background of ``noise'' for the vary rare microlensing events.  He
did not know that the atmospheric seeing is blurring stellar images
in a different way every night.  He did not know many other things.
The bottom line was: he did not know the project could not be done,
and so he tried to persuade me, a seasoned astronomer, that
the project was feasible.  Well, I knew better, and so he failed.

But Dave Bennet was a stubborn fellow.  He went off to California and
tried to persuade Charles Alcock, who was a physicist as well as an
astronomer.  And the interdisciplinary gap was bridged for the second
time.  Charles Alcock gave a colloquium in Berkeley, outlining the
project: the search for dark matter by means of monitoring the brightness
of a few million stars in the nearby galaxy, the Large Magellanic Cloud.
The response was enthusiastic.  So much so that Jim Rich, visiting
from Sacley, France, immediately called his collaborators and adviced them to
start their own experiment.  And so the first two microlens searches were
founded: the American-Australian collaboration of 18 participants,
called MACHO (MAssive Compact Halo Objects), and the French collaboration
of 28 participants, called EROS (Exp\'erience de Recherche d'Objects Sombres).
Both teams had some astronomers among them but they were really dominated
by the particle physicists.

I did not know about all these developments at the time.  I was dreaming
of having a personal 1 meter telescope.  First I thought that funds were
the limiting factor.  Pretty soon I realized that even though I had no
funds there was another, much more limiting factor: I had no instrumental
skills.  But I was lucky.  A friend pointed out to me that there was a
truly outstanding instrumentalist at the Warsaw University Observatory,
Andrzej Udalski.  Being well connected to Warsaw (I lived there till
1981) I contacted Andrzej Udalski, and pretty soon we were dreaming
together.  Now all we needed were the funds, and naturally a good
observing site.  I can no longer recall how the Las Campanas Observatory
(operated by the Carnegie Institution of Washington) came about.  Perhaps
it was the presence of our old friend from Warsaw there: Wojtek Krzeminski
worked for Carnegie since 1982.  He lived in Chile, and he knew everything
about Chile.  One way or another, one day I got a phone call from
George Preston from Pasadena, California, the headquarters of the
Carnegie Observatories.  George Preston proposed some form of collaboration
between Carnegie, Warsaw and Princeton.  I think it was rather vague
initially, but pretty soon two somewhat independent, somewhat related
projects took off.

One project was a long term and fairly ambitious.  The idea was that
the Warsaw University Observatory was to build their own 1-meter class
telescope at the Las Campanas site.  The telescope was to be dedicated
to long term massive photometric searches of variable stars of all kinds,
among them the microlensing events.  Carnegie would contribute the site,
Princeton - the operating expenses.  The telescope would be managed
from Warsaw.  Marcin Kubiak, the director of the Warsaw University
Observatory begun the fund raising effort.  It did no harm to have a
friend astronomer Robert Glebocki as the Minister of Higher Education
in the first post-communist government in Poland.  But the telescope
could be operating no sooner than 1995.

In the mean time another, more modest project took shape.  A bunch of
8 astronomers from Warsaw, Pasadena and Princeton applied for about
60 nights in the 1992 season on the Swope 1-meter telescope, the oldest
one at Las Campanas.  We were to monitor about 1 million stars in the
galactic bulge, close to the galactic center.  The idea was to check
if we could detect microlensing events where they were guaranteed to
be present, as there are many ordinary stars in the galactic disk,
between us and the million stars located at the galactic bulge.  As a
by product we were going to find thousands of variable stars, a real
treasure for us, as our team was all made of professional variable
stars observers.  It is worth pointing out that the same variable
stars were mostly noise from the point of view of particle physicists
in their quest for dark matter.  We were lucky: we got between 70 and
80 nights on the telescope for each of the three observing seasons:
1992, 93, 94, and we have a good chance to be awarded a similar
amount in 1995, at which time the transition will take place to the
new instrument, which grew up to become a 1.3 meter telescope.

What we needed now was the name - so we could be recognizable as a team.
Our first clumsy attempts in a search for a good acronym made George
Preston go non-linear in outrage.  He was an old friend of mine - I worked
for him in 1962/63 as an observer at Lick Observatory in California,
and he is a man of many talents.  In just one afternoon he created a
list of about 10 acronyms, all great, and we e-mailed the list to Warsaw -
let them decide which should be our name.  And so the OGLE (Optical
Gravitational Lensing Experiment) came to be.

The next major event was the early morning phone call from Charles
Alcock.  It was September 1993, and I was in Warsaw for a few days.
I was told the MACHO had their first microlensing event, or the
candidate event as it was cautiously named.  That was fascinating!
I was very excited.  It was good, as my 1986 idea was sound.  It
was not so good as OGLE had no event yet.  It turned out that MACHO
had their event for some weeks, but they were waiting for more data before
going public.  However, they were notified by Jim Rich that EROS had
two events to be announced at something that sounded to me like a meeting
of underground physics in Italy.  One way or another MACHO decided
to make their announcement at the same meeting.  Unfortunately,
OGLE had nothing to announce, and even if we had we were in no way
connected to the underground physics.

Within a week of my return to Princeton I got a phone call from
Marcin Kubiak in Warsaw: the first OGLE event has been found!
By that time the media was full of excitement about the discovery of
dark matter.  Unfortunately we were one week too late, and we missed
the boat.  We also knew that MACHO and EROS submitted their
announcement papers to Nature which is well known for a very short
publication cycle.   There was no way we could make it for the
same issue.  On top of that I had very bad personal relations with
Nature, and this would not help with a speedy publication.  Fortunately,
Marcin Kubiak happened to be the editor of Acta Astronomica, a
quarterly journal.  Normally, to wait 3 months for a publication
would be of no use in this high speed month of October 1993.
By some incredible twist of luck the last issue had been just sent
to the publisher.  Kubiak called and asked: "can you hold printing
for two days?".  The publisher agreed, and the discovery paper was
written and recycled a few times between all the authors
scattered on two or three continents, all in 48 hours - the marvels of
computer networking.  As it turned out the issue of Acta Astronomica
with the OGLE paper arrived to some libraries a few days ahead of Nature
with the papers of the two competing groups.

This irritated some.  I got angry e-mails from a few EROS
participants: why did we not quote their work presented in Italy
at a conference?   My answer was simple.  We have never received
any information from EROS directly, no preprint, no e-mail.  The
only tangible evidence of their discovery we had from the article in
the Time magazine.  I wanted to quote that article in our Acta
Astronomica paper, but I was outvoted by all other OGLE members - they
felt quoting Time would be too frivolous.  I am still sorry we did not
do it.   When the issue of Nature finally arrived I could read the
two papers.  I discovered that the very existence of OGLE was not
even mentioned in the EROS paper,   That made me feel better about
not quoting Time about EROS.  In any case, some good came from the
incident.  Our purely astronomical team has been noticed by the mostly
elementary particle team.

There were more surprises as time went by.  Many people, including
myself, took it for granted that while the OGLE event was most likely
caused by an ordinary star somewhere in the galactic disk, the MACHO
and the EROS events were caused by brown dwarfs, which presumably
made up the dark matter.  Not everybody agreed.  Some, including
Jeremiah Ostriker at Princeton, were pointing out that the time scales
of all events were about the same, a few weeks, hence they may all
be caused by the same type of objects, perhaps just ordinary stars
located somewhere between us and the Magellanic Clouds.  The confusion
grew.  At the same time OGLE kept finding new events towards the
galactic bulge.  We had a feeling there were too many of them.
When I was again in Warsaw in January 1994 a student, Marcin Kiraga,
pointed out to me, that even though the geometric location of the
galactic disk stars makes them more efficient for gravitational
lensing, the shear number of the galactic bulge stars more than
outweights their geometrical disadvantage.  We quickly wrote
a joint paper together presenting quantitative evidence that the
majority of the bulge events are most likely due to other galactic bulge
stars acting as lenses (Kiraga and Paczy\'nski 1994).

No new official reports were available from the MACHO and EROS teams,
but there was a general feeling that the rate of events towards the
Magellanic Clouds was much too low to account for all the dark matter.
Perhaps those few events could really be accounted for by the ordinary
stars in our galaxy?  In the spring of 1994, back in Princeton, I received
two papers on the same day.  One was published in the Astrophysical Journal,
and its title was: ``Microlensing Events: Thin Disk, Thick Disk, or Halo?''
(Gould et al. 1994).  The second was a paper from Nature to referee.
An unknown post-doc Kailash Sahu (1994)
(from Instituto de Astrofisica de Canarias at Tenerife, Canary Islands)
pointed out that the lenses towards the Magellanic
Cloud might be the ordinary stars in the Magellanic Cloud itself.  I was
struck with the simplicity of the idea: it  was basically the same as
Kiraga's, just applied to the Magellanic Cloud rather than the galactic
bulge.  Of course the two men came up with their insights independently.
They also shared their status - both were unknown outsiders.  All the
prominent people working in the field missed their point.  I missed it
twice.

By June 1994 I was at the Institut d'Astrophysique in Paris, and
OGLE has come up with their first estimate of the so called optical
depth to gravitational microlensing towards the galactic bulge stars
based on 9 events found in a systematic search in the 1992 and 1993 data
(Udalski et al. 1994a,b).  The optical depth is a number proportional to
the probability of finding the lensing events.  Theoretically it was
supposed to be between 0.5 and 1.0 parts in a million, it turned out
to be $ 3.3 \pm 1.2 $.   MACHO also came up with a similar estimate
based on their 4 events (Alcock et al. 1994).  This was a major surprise:
what was going on?  My tentative speculation is that the excess may be
accounted for if the galactic bulge is in fact a galactic bar, with its
long axis pointing more or less towards us: the same number of stars
can act as lenses much more efficiently (Paczy\'nski et al. 1994).  About
50\% of all spiral galaxies are known to have bars at their centers, and
de Vaucoulers (1964) presented compelling evidence for the bar in our own
galaxy.  Now this is a common knowledge among the few specialists,
but it is practically unknown to general astronomers or physicists.
So, in 1986 I rediscovered gravitational microlensing, and in 1994
the OGLE search has rediscovered the galactic bar.  I seem to be
valways a few decades late.

As for the dark matter - it is still a mystery.  There is no doubt that
the recent experiments have demonstrated that the technology works:
the monitoring of millions of stars for a few years, and the detection of
the few microlening events has been done.  Even the detection of the first
double lens has been reported (Udalski et al. 1994c).  The results are
already useful for the studies of galactic structure.  Perhaps in the
future the same technique will lead to the discovery of dark matter, but
we have no clear evidence that this has already happened.

\centerline{-----------------}

{}From the beginning of 1994 season the OGLE project is capable of near real
time data processing (Paczy\'nski 1994). The new computer system automatically
signals the events while they are on the rise, making it possible to carry out
photometric and/or spectroscopic follow-up observations. The observers who
would like to be notified about the on-going events should send their request
to A.~Udalski (udalski@sirius.astrouw.edu.pl).

The photometry of the OGLE microlensing events, their finding charts,
the updated OGLE status report, including more information
about the ``early warning system'', can be found over Internet at
``sirius.astrouw.edu.pl'' host (148.81.8.1), using the ``anonymous ftp''
service (directory ``ogle'', files ``README'', ``ogle.status'',
``early.warning''). The file ``ogle.status''
contains the latest news and references to all OGLE related
papers.  PostScript files of some papers, including
Udalski et al. (1994b), are also available.
The OGLE results are also available over
``World Wide Web'': ``http://www.astrouw.edu.pl/''.

{\bf References:}
\vskip 0.5cm
\item{}
Alcock, Ch. et al. 1993, Nature, 365, 621.
\item{}
Alcock, Ch. et al. 1994, Ap.J., submitted.
\item{}
Aubourg, E. et al. 1993, Nature, 365, 623.
\item{}
Blitz, L. 1993, in AIP Conference Proceedings 278, p. 98.
\item{}
de Vaucouleurs, G.  1964, IAU Symp. 20, p. 195.
\item{}
Gould, A., Miralda-Escud\'e, J. and Bahcall, J. N. 1994, Ap.J., 423, L105.
\item{}
Hammond, R. 1994, Matter of Gravity, No. 3, p. 25.
\item{}
Kiraga, M., and Paczy\'nski, B. 1994, Ap.J., 430, L101.
\item{}
Paczy\'nski, B. 1986, Ap.J., 304, 1.
\item{}
Paczy\'nski, B. 1994, IAU Circular No. 5997.
\item{}
Paczy\'nski, B., et al.  1994, Ap.J., submitted.
\item{}
Sahu, K.  1994, Nature, 370, 275.
\item{}
Udalski, A., et al.  1993, Acta Astronomica, 43, 289.
\item{}
Udalski, A., et al.  1994a, Ap.J., 426, L69.
\item{}
Udalski, A., et al.  1994b, Acta Astronomica, 44, 165.
\item{}
Udalski, A., et al.  1994c, Ap.J., submitted.

\vfill
\eject
\centerline{\bf Laboratory Gravity:  The G Mystery, and Intrinsic Spin
Experiments.}
\medskip
\centerline{Riley Newman, University of California, Irvine}
\centerline{rdnewman@uci.edu}
\bigskip
\bigskip
\medskip

\noindent{\bf $\bullet $ Disconcerting G News:}  The accepted \lq\lq CODATA"
value for G, based on the 1982 measurement of Luther and Towler
[1], carries an uncertainty of 128 ppm.  In the last newsletter, I
reported a startling result from the PTB (German NIST equivalent)
group: G a whopping {\bf 0.6\% higher} than CODATA.  This
was officially announced at the NIST-sponsored \lq\lq CPEM"
conference in Boulder this summer [2]. At the same conference the
Wuppertal group [3] reported a G consistent with CODATA, but the
Measurement Standards Laboratory of New Zealand [4] presented a G
{\bf 0.1\% lower} than CODATA!?!!

Not at CPEM, but just published in English translation [5], are
the results of a Russian group that finds inexplicable temporal
variations at a level of about 0.7\% in their measured G (for mass
separations about 18 cm;  for mass separations about 6 cm the
variations appear smaller).  The Russian group is thoroughly
puzzled, and seriously considers the possibility of a real time
variation in an effective G.

Meanwhile, Luther with C. Bagley are developing an instrument that
will use and compare both of the two customary approaches to the
use of a torsion balance for determining G:  measurement of
equilibrium angle, and of torsional period.  This will address
speculation that the period measurement approach might have
undiscovered systematic effects linked to torsion fiber
properties.

The following table summarizes the disconcerting recent results:
\medskip

\hrule
\settabs\+Codata value for G&mercury support for
&6.6666666666&assigne undertainty&-101000&deviations\cr
\+&&&&{\bf G - G $_{CODATA}$}\cr
\+  Source       &  Method  & G $\cdot 10^{11}$   & Assigned   &
ppm  & standard  \cr
\+ &  &  &  uncertainty(ppm)   &  & deviations \cr
\smallskip\hrule
\+CODATA value$^1$  &  fiber torsion   & 6.6726  &  128  &  $\equiv
0$  &  $\equiv 0$ \cr
\+         &     balance, period \cr
\smallskip\hrule
\+PTB$^2$  &  mercury supported   &  6.7154  &  ~83  &  + 6400  &  +
42.0  \cr
\+   &  torsion balance, \cr
\+   &   static \cr
\smallskip\hrule
\+Wuppertal$^3$  &  dual pendulums,   &  6.6719  &  117  &  - 105  &
- 0.6\cr
\+    & static \cr
\smallskip\hrule

\+New Zealand$^4$  &   fiber torsion   &  6.6656  &  ~95  &  - 1050
&  - 6.6  \cr
\+   &  balance, static \cr
\smallskip\hrule

\+Russia$^5$\cr

\+~~~~~ @18 cm  &  fiber torsion  &  6.6655  &  (0.7\% temporal  &
- 1060  \cr
\+& balance, period  &  & variation)  \cr
\+~~~~~ @ 6 cm  &  & 6.6771  &  &  + 670\cr
\smallskip\hrule

\noindent{\bf $\bullet$ Intrinsic Spin Experiments}  The possibility of a
special dependence of the gravitational interaction on intrinsic
spin has provoked experimentalists for many years.  Several new
results put greatly improved limits on such behavior.

\noindent{\bf Spin-spin interactions}  An elegant new experiment
by Wei-Tou Ni and collaborators [6] puts a convincing limit $(1.2
\pm 2.0) \times 10^{-14}$ on the ratio of the strength of an
anomalous electron spin-spin interaction (of the same form as the
magnetic interaction) to that of the magnetic interaction.  Some
background is required here:

Early tests for such interactions used torsion balances [7,8].
Two Russian experiments [9,10] applied an ingenious and
potentially far more sensitive technique, based on the idea that
an anomalous interaction from a spin-polarized source should
penetrate a superconducting magnetic shield and induce a spin
polarization in a shielded ferromagnetic sample.  The induced
polarization would produce a magnetic flux, to be sensed by a
SQUID.  But there is a serious concern here, raised by Ritter [7]:
might the ferromagnetic domains be too \lq\lq sticky" to respond
to the extremely small aligning torques associated with the
sought-for anomalous interaction?  This concern is avoided in Ni's
experiment, which uses the basic technique pioneered by the
Russian groups, but substitutes a high susceptibility
{\bf paramagnetic} sample for a ferromagnetic one.

\noindent{\bf $\sigma \cdot \hat{r}$ (monopole-dipole)
interactions}  Does the gravitational interaction of a polarized
particle depend on whether its spin points away from or towards
the source of the force?  For ranges greater than about one meter,
the best limits on this (for nucleons) come from recent NMR
experiments [11,12].  Venema et al. conclude (invoking a minimal
bit of nuclear modeling) that the energy difference between a
spin-up and spin-down neutron in the Earth's gravitational field
is less than $2.1 \times 10^{-20}$ eV (compared to its total
gravitational energy of about 0.7 eV).  Limits on the fractional
contribution of spin to total energy get rapidly weaker as shorter
interaction ranges are assumed; below 1 meter the best limits (for
electrons) have come from torsion balance experiments of Ni et al.
[13] and Ritter et al. [14], giving comparable limits which may be
expressed: the spin-dependent part of the force on an electron
about 4 cm from an attracting mass is less than about 5000 times
the total gravitational force due to that mass.

But the same clever trick that the Russian groups and Ni have
applied to spin-spin interaction searches has now been applied by
Ni and collaborators to search for an electron $\sigma \cdot
\hat{r}$ interaction:  a paramagnetic sample within a
superconducting shield is examined for induced magnetism as an
unpolarized mass is rotated around the outside of the shield.
Results, announced by Ni at MG7 this summer, improve limits on
such an interaction for ranges between a few centimeters and one
meter by nearly two orders of magnitude.

\noindent{\bf Does intrinsic spin recognize an anisotropy in
space?}  A recent torsion balance experiment by Wang, Ni, and Pang
[15] puts a limit of $3.5 \times 10^{-18}$ eV on the energy
splitting of the spin states of an electron relative to the
spacial directions explored as the earth turns.  A similar
experiment, aimed especially at possible effects associated with
galactic dark matter, is being conducted by Ritter [16].

\noindent[1]  G.G. Luther and W.R. Towler, Phys. Rev. Lett. 48, 121 (1982)

\noindent[2]  W. Michaelis, H. Haars, and R. Augustin*, Physikalisch-
Technische Bundesanstalt (PTB), Postfach 3345, 38023,
Braunschweig, Germany.  FAX 49531 5924015

\noindent[3]  H. Walesch, H. Meyer, H. Piel, and J. Schurr*, Bergische
Universit$\ddot{a}$t Wuppertal, Gaussstrasse 20, D-42119,
Wuppertal, Germany\hfil\break E-mail: meyer@wpvx11.physik.uni-wuppertal.de

\noindent[4]  Mark Fitzgerald and Tim Armstrong*, Measurement Standards
Laboratory of New Zealand.  E-mail: t.armstrong@irl.cri.nz

\noindent[5]  V.P. Izmailov, O.V. Karagioz, V.A. Kuznetsov, V.N. Mel'nikov,
and A.E. Roslyakov, Measurement Techniques {\bf 36}, no. 10, 1993.
Oleg Karagioz: amcor@tribo.ecosrv.msk.su

\noindent[6]  Wei-Tou Ni et al, Physica B, {\bf 194-196}, 153 (1994), Phys.
Rev. Lett. 71, 3247 (1993).

\noindent[7]  R.C. Ritter et al, Phys. Rev. D{\bf 42}, 977 (1990).

\noindent[8]  S.S. Pan, W.T. Ni, and S.C. Chen, Mod. Phys. Lett. A {\bf 7},
1287 (1992).

\noindent[9]  P.V. Vorobyov and Ya. I. Gitarts, Phys. Lett. B {\bf 208},
146 (1988).

\noindent[10] V.F. Bobrakov et al., JETP Lett.{\bf 53}, 294 (1991).

\noindent[11] D.J. Wineland et al., Phys. Rev. Lett. {\bf 67}, 1735 (1991).

\noindent[12] B.J. Venema et al., Phys. Rev. Lett. {\bf 68}, 135, (1992).

\noindent[13] T-H Jen et al., MG6, Kyoto Japan.

\noindent[14] R.C. Ritter, L.I. Winkler, and G.T. Gillies, Phys. Rev. Lett.
{\bf 70}, 701 (1993).

\noindent[15] S-L Wang, W-T Ni, and S-S Pan, Modern Physics Letters A {\bf
8}, 3715 (1993).

\noindent[16] R.C. Ritter, L.I. Winkler, G.T. Gillies, Proceedings of the
January 1994 Moriond Workshop.

* Results 2,3,4 were presented at the Conference on Precision
Electromagnetic Measurements (CPEM), Boulder Colorado, 27 June - 1
July, 1994.  Result 3 was presented also at MG7 at Stanford; the
value for G quoted here was that given at MG7, where I have added
formal and systematic error estimates in quadrature.

\vfill
\eject

{\centerline {\bf LIGO Project update}}
\medskip
{\centerline {Stan Whitcomb, Caltech}}
\centerline{stan@ligo.caltech.edu}
\bigskip
\bigskip
\medskip

A major step for LIGO took place this spring when construction of the
first of the two 4-km LIGO facilities began at Hanford, Washington.
The official ground-breaking ceremony was held on July 6, against a
backdrop of earthmoving equipment already at work leveling the site and
preparing the soil for the foundation.  This inital earthwork is
expected to be completed later this fall.

At the other LIGO site in Livingston, louisiana, the purchase of the
land by Louisiana State University (LSU)  and a
lease transferring the site to NSF are in the final signature stage.
With the cooperation of the original landowner and LSU, we were able to
arrange for the clearing of the site, begin the on-site geotechnical
investigations, and apply for the necessary wetlands permit from the
Army Corps of Engineers, in advance of the final land transfer.
Progress on these items should help speed the start of construction at
the Louisiana site.

The engineering design of the beam tubes which span the full
4km arms of the two facilities has been completed.  A qualification
test of the design using full diameter tube segments and the field
assembly procedures is being carried out by our design contractor,
Chicago Bridge and Iron, Inc.  The successful completion of this
test, expected near the end of this year, will clear the way to
begin fabrication of the required 16 km of beam tube.  The other
major engineering design contracts are for the building and site
design and for the remainder of the vacuum system.  The Request for
Proposals (RFP) for the  building and site design has been released to
qualified Architect-Engineering firms, and the RFP for the remainder
vacuum system is in preparation.

As exciting as the beginning of construction is, some of the best news
has come from the R\&D work.  The reconstructed 40 m interferometer at
Caltech, called "Mark II" to distinguish it from its predecessor, has
come into full operation now.  This new version of the interferometer
is housed in a new vacuum system, one which gives us much more room for
testing concepts and hardware for the full-scale LIGO interferometers.
The most significant change to the interferometer itself, namely, the
replacement of the seismic isolation stacks, has resulted in an
improvement in the low frequency performance of the interferometer by
as much as a factor of 100.  The next major change to the 40 m
interferometer (currently underway) is the replacement of the old test
masses with new ones which have the required supermirror coating
deposited on on a polished face of the test mass itself, eliminating
the need to have a mechanical attachment of a separate mirror to the
test mass.  The preliminary indications are that this change will
reduce the amount of thermal noise in the interferometer and give yet
another improvement in performance.

A major effort to demonstrate the optical phase sensitivity required
for the LIGO interferometers is underway at MIT.  A 5 m interferometer
is being built with seismically-isolated, suspended mirrors.  This
interferometer is designed to operate with the same laser power
incident on its beamsplitter as the full-scale LIGO interferometer.
Thus, it should have the same shot noise  (measured as an uncertainty
in optical phase) as the LIGO interferometers.  In this way, it will
test for any other optical noise sources and verify that they can be
controlled with adequate precision.  Together, the results from the 5 m
and 40 m interferometer should provide a firm foundation on which to
base the LIGO interferometer design.

Many of the other R\&D activities have less readily explained results
(for the nonexpert), though the work is not necessarily less important
or less difficult! We have made steady progress in understanding the
complexity of the alignment systems required, in establishing the
capability of industry to fabricate the high precision LIGO optics, in
developing the input optics which control and stabilize the laser beam,
and in modeling the various aspects of the interferometer. All of these
efforts are essential to a success project.

The past six months have also seen a reorganization of the project to
enhance the project management.  Barry barish has been appointed as
Principal Investigator, and Gary Sanders has been recruited from Los
Alamos to become Project Manager.  The new management plans will be
reviewed by NSF in September along with an updated cost estimate.
Success at this review (which we expect) will give us the go-ahead for
the increasing level of effort required to complete LIGO on schedule.

\vfil
\eject
\centerline{\bf PASCOS '94}
\medskip
\centerline{Peter Saulson, Syracuse University}
\centerline{saulson@suhep.phy.syr.edu}
\bigskip
\bigskip
\medskip

	PASCOS '94, the 4th meeting in the series called Particles,
Strings, and Cosmology, was held at Syracuse University from May 19
through May 24.  Remarkably fine weather surprised the organizers and
delighted the participants when the latter weren't glued to their
chairs for a stimulating set of invited talks. Although at times the
diverse program seemed about to demonstrate that physicists form a set
of communities separated by a common language, great efforts were made
to bridge our cultural differences. An idiosyncratic listing of
highlights of the meeting follows.

	Two excellent talks on the solar neutrino opportunity
("problem" is too negative a word) were presented by Ettore Fiorini,
representing the experimental community, and by John Bahcall, who
tried to make his talk look like an experimenter's by displaying
histograms of the results of one thousand different solar models. Both
agreed that the glass of neutrinos was half full, the detection of
solar neutrinos at roughly the right rate demonstrating that the Sun
shines by fusion of hydrogen into helium. The persistent deficit in
neutrino fluxes (consistent results from SAGE and GALLEX joining the
list with Kamiokande and the pioneering Homestake Mine Experiment) is
called an opportunity to learn new physics. The MSW effect is the
odds-on favorite.

	An upbeat presentation on the status of the next hard
experiment, LIGO, was given by Deputy Director Stan Whitcomb. Site
work has begun at Hanford, and a contract for engineering design of
the beam tubes has been let. The schedule calls for the first site to
be available for interferometer installation in 1998, and for the
first coincidence run between the two sites around 2000.

	Astrophysical cosmology was the subject of a set of excellent
talks.  Ned Wright gave a very clear summary of the results from COBE,
and of the limits set by those observations on standard and
not-so-standard cosmologies. The inflationary paradigm gets a clean
bill of health. The topic was picked up by David Spergel in his talk
entitled "Examining the Wreckage", in which he acknowledged that "many
of our theories are stretched to the breaking point."  Understanding
the rich data base now accumulating on large scale structure,
especially the comparison between the COBE map's power spectrum and
the correlations between galaxies revealed by the ever-improving
surveys, seems to be difficult in the context of any model simple
enough to have been considered beautiful heretofore. Multi-component
dark matter, non-zero cosmological constant, "tilted" fluctuation
power spectra, or a Hubble constant even lower than Sandage's are
among the bizarre alternatives taken seriously now. Spergel and Joshua
Frieman gave summaries of the large number of new observational
efforts that will, with luck, make theorists' work even harder in the
coming years.

	A beautiful description of the status of searches for Massive
Compact Halo Objects (MACHOs) was presented by David Bennett. He
showed convincing evidence for several microlensing events, but
perhaps the highlight of his talk was the evidence for the link
between experiment acronyms and national character: MACHO
(American/Austrialian), EROS (French), and OGLE (Polish).  The
international effort seems to have established microlensing as a real
phenomenon, but at a disappointingly low rate. The dark matter problem
is not solved yet.

	J.V. Narlikar gave a classic talk on the Quasi Steady State
Cosmology, which can be adjusted to look arbitrarily similar to the
Hot Big Bang Cosmology. Perhaps the steadiest thing about the Steady
State theory is its adherents' determination to make some version of
it work.

	A session that included talks by Andy Strominger, Lenny
Susskind, and Abhay Ashtekar was a remarkable example of cultural
convergence. The first two  devoted their
talks to black holes.  Strominger spent a portion of his talk
explaining Penrose diagrams to the uninitiated, while Susskind (who
prefers Kruskal diagrams) was effusive in his praise of the membrane
paradigm. Each was selling a different "stringy" solution to the
information problem in black holes. (In response to a question,
Strominger admitted to having swept under the rug a naked singularity,
but said he didn't think it was important, since it was "only at one
point.") Ashtekar, in the meantime, described progress in
relativity-inspired quantum gravity research with a distinct emphasis
on the analogies to string theory, albeit formal ones for the most
part.

	The prize for the melding of cultures must be awarded, though,
to Mark Bowick, whose outstanding talk on studies of the evolution of
defects in liquid crystals managed to be simultaneously about early
universe cosmology, high energy theory, and experimental condensed
matter physics.

	Among the highlights from the strictly particle physics side
of the meeting was the detailed presentation by Henry Frisch of the
evidence for (don't say "discovery of") the existence of the top
quark. Perhaps in Stockholm they should give out evidence for the
Nobel Prize at the 2 to 3 sigma level.  Riccardo Barbieri gave a pep
talk for the prospects of supersymmetry, an idea that almost rivals
gravitational waves for its longevity without an experimental
discovery.

	The conference was characterized by many lively exhanges
between speakers and questioners. Some nearly-verbatim quotes: "It is
remarkable that you got money when you didn't know what you were
doing," "That isn't physics, it's meaningless," "I proved your idea
was incorrect in the referree's report I wrote on your last paper,"
"We won't settle this here, we'll settle it later [with gesture toward
the door]", and "Thank you for giving my talk."

	After the banquet on May 23, science journalist Gary Taubes
regaled diners with the dirty details surrounding the cancellation of
the SSC. But perhaps the clearest summary of the intellectual
enterprise represented at the meeting came in entirely non-verbal form
on the previous evening. Two gifted women, Roxanne Kamayani Gupta and
Amita Dutta, gave a recital of classical Indian dance. They performed
on a stage graced with a statue of Shiva Nataraja, the Lord of the
Dance, whose connection to physics, we were reminded, was made clear
by Fritjof Capra. Dancing in two highly elaborate and remarkably
distinct styles, without speaking a (non-Sanskrit) word, the two
managed to embody the quest for beauty that unites us all.

\vfil
\eject

\centerline{\bf The Vienna meeting}
\medskip
\centerline{P. C. Aichelburg, R. Beig}
\centerline{fwagner@pap.univie.ac.at}
\bigskip
\bigskip
\medskip

{}From July 25 to July 29, 1994, a conference on Mathematical
Relativity was held at the Erwin Schr\"odinger Institute for
Mathematical Physics (ESI) in Vienna, organized by P. C. Aichelburg
and R. Beig. The conference was attended by 90 people, roughly one
third of whom were also participants in the ESI workshop on the same
topic running from 1st July through 'til 16th September 1994. The
talks at the conference focussed on one hand on soliton type solutions of
the Einstein equations coupled to nonlinear matter fields, on the other
hand on global issues in classical General Relativity in general.

The one hour talks were:
R. Bartnik - The Non-Abelian Einstein-Kaluza-Klein System,
P. Bizo\'n - On the No-Hair Conjecture,
D. Brill - Testing Cosmic Censorship with Black Hole Collisions,
M. Choptuik - Critical Phenomena in Gravitational Collapse,
Y. Choquet-Bruhat - Non Abelian Relativistic Fluids,
P. Chru\'sciel - Strong Cosmic Censorship in Locally Homogeneous
 Spacetimes,
H. Friedrich - Boundary Conditions for Anti-de-Sitter Type
 Spacetimes,
G. Gibbons - Gravitating Solitons,
W. Israel - Effect on Radiative Wave Tails on Black Hole Interiors,
D. Maison - Analytical and Numerical Methods for Einstein-Yang-Mills
 and Related Systems,
V. Moncrief - The Reduction of Einstein's Equations in 3+1
 Dimensions and the Analogue of Teichm\"uller Space,
K. Newman - The Structure of Conformal Singularities,
N. \'O  Murchadha - Spherical Gravitational Collapse,
G. Rein - On the Spherically Symmetric Vlasov-Einstein System,
A. Rendall - Crushing Singularities in Spherically Symmetric
 Cosmological Models,
B. Schmidt - The Newtonian limit of Einstein's Theory of Gravitation,
N. Straumann - On the Einstein-Yang-Mills System for Arbitrary Gauge,
R. Wald - Classical Thermodynamics of Black Holes in Arbitrary,
Lagrangian Theories of Gravity Coupled to Matter.

The half-hour talks were:
P. Brady - Critical Behaviour in Self-Similar Scalar Field Collapse,
M. Iriondo - Existence and Regularity of CMC Hypersurfaces in
Asymptotically Flat Sapcetimes,
H. Pfister - Dirichlet Problem for the Stationary Einstein Equations
with Applications to Stability Limits of Rotating Stars,
B. Temple - An Astrophysical Shock-Wave Solution of the Einstein
Equations Modelling an Explosion,
G. Weinstein - N-black Hole Stationary Axially Symmetric Solutions
of the Einstein-Maxwell Equations.

An ESI preprint containing the abstracts of these lectures should be
available by September and can be obtained via `anonymous ftp',
`gopher' or `WWW' : FTP.ESI.AC.AT
\vfil
\eject

\centerline{\bf The Pitt Binary Black Hole Grand Challenge Meeting}
\medskip
\centerline{Jeff Winicour, Univesity of Pittsburgh}
\centerline{jeff@artemis.phyast.pitt.edu}
\bigskip
\bigskip
\medskip

A meeting of the BINARY BLACK HOLE GRAND CHALLENGE ALLIANCE was held at
the University of Pittsburgh on May 5-6, 1994. This was the first
meeting planned since the project was funded so that the major emphasis
was placed on organizational matters and background presentations for
the benefit of the Advisory Committee, which Kip Thorne had just agreed
to head. Two meetings are anticipated each year, which will rotate
between the sites of the eight participating universities: Cornell,
Illinois, North Carolina, Northwestern, Penn State, Pitt, Syracuse and
Texas.  One of these meetings will be in the format of an open
numerical relativity conference, as held at Penn State last Fall, and
the other will be focussed on keeping the Alliance on track. The Pitt
meeting gave the co-PI's, research scientists, graduate students and
other collaborators on the Grand Challenge a chance to meet, show their
progress and what they had to offer, develop further collaborations,
identify problems and methods of attack and give feedback to overall
and individual plans. The Proceedings, comprising approximately 200
pages of handouts and xeroxes of transparencies, may be ordered by
sending a check for \$20, made out to the University of Pittsburgh, to
Dr. Roberto Gomez, Department of Physics and Astronomy, University of
Pittsburgh, Pittsburgh, PA 15260.

The main goal of the project is to supply a catalog of gravitational
waveforms for the inspiral and coalescence of two black holes. Richard
Matzner, the project leader, gave an overview of the necessary
subtasks, their schedules for completion, how they fit together and the
collaborations that had been formed among the groups. The programs,
timelines and progress of the relativity groups were presented by
co-PI's S. Shapiro and S.  Teukolsky (Cornell), L. Smarr and E. Seidel
(Illinos), C. Evans  (North Carolina), S. Finn (Northwestern), P.
Laguna (Penn State), J. Winicour (Pitt), and R. Matzner (Texas). The
issues bearing on the computer science groups were detailed by co-PI's
P. Saylor and F. Saied (Illinois), G.  Fox (Syracuse) and J. Browne
(Texas, unable to attend due to illness but in a written version).  W.
Rheinboldt (Pitt) gave a survey of DAE's (constrained systems of
differential algebraic equations), which are used in the computational
approach of the Illinois group. Also, E. Newman (Pitt) surveyed the eth
formalism, now being developed in finite difference form by the Pitt
group to allow use of spherical coordinates in 3D tensorial problems.
The trends in supercomputing at the national centers and the current
transition from vector to parallel and heterogeneous processing were
outlined by L. Smarr (NCSA), G. Fox (NPAC) and R. Roskies (PSC).

The main relativity topics discussed were the development of 3D codes;
finding and tracking the apparent and real horizons, which enter into
the interior boundary conditions for a Cauchy evolution; matching the
exterior Cauchy boundary to a characteristic evolution in order to
supply an outgoing radiation boundary condition and propagate the
radiation to scri; the physics of the initial data set; and numerical
issues such as elliptic solvers, coordinate systems and adaptive mesh
refinement. The horizon problem, involving the trouser shaped merger of
the two black holes, is potentially the most difficult of these but
remarkable progress, in the case of head-on collisions, was presented
by the Cornell and the Illinois groups (in collaboration with W. Suen
of Washington U.). They are able to track the jump in the apparent
horizon and follow it for sufficient time for it to settle down close
to the event horizon. A late time slice of the apparent horizon is then
used to reconstruct the actual horizon by tracing back a null
hypersurface.

The computer science aspect of the project is important for developing
a backbone structure that will allow the various groups to interface
codes and data and to spearhead the adoption of High Performance
Fortran standards which will enhance use of the emerging generation of
teraflop machines. Roughly half the meeting was devoted to these
issues. This has resulted in a proposal for a uniform Fortran 90 data
structure, which can be examined using mosaic/www with URL

http://www.npac.syr.edu/NPAC1/PUB/haupt/bbh.html

\noindent (in the 'data
structures' section). This standard, which all groups will adopt by
Fall 1994, is the first step toward developing a software toolkit for
the project.

The meeting concluded with an administrative session with the freshly
appointed Chairman of the Advisory Committee. Kip reported estimates of
signal to noise ratio in the tens (for advanced LIGO) from inspiraling
black holes of tens of solar masses at distances of 500Mpc. This
supplies practical feedback for the minimal quality of Black-Hole
waveforms that should constitute the Alliance catalog.  Kip also
related the impression: It seems that every group wants to do
everything in order to solve the waveform problem. This reflected the
energy and skill of the groups as well as the individualistic tradition
of general relativity. The meeting had made progress in setting the
Alliance on a cooperative course in crucial areas where collaboration
would speed progress and avoid duplication of effort.

\vfil
\eject

\centerline{\bf International Symposium on Experimental
Gravitation, Nathiagali, Pakistan, 1993}
\medskip
\centerline{Munawar Karim, St. John Fisher College, Rochester NY}
\centerline{karim@sjfc.edu}
\bigskip
\bigskip
\medskip

The main purpose of the symposium was to provide a forum for
physicists engaged in designing/performing/thinking about experiments in
gravitational physics. In order to allow maximum participation apart from
invited talks there were few contributed papers which could be read to all
participants without any parallel sessions.
There were 75 participants from 17 countries from five contintents. The
Proceedings have been published by the Institute of Physics Publishing (UK)
and are available from them. Their address is: Techno House, Redcliffe Way,
Bristol BS1 6NX, UK.  The invited papers are reprinted in  a supplementary
issue of the June 1994 issue of the journal Classical and Quantum Gravity.
The subjects covered in the Symposium described experiments to test the
predictions of the general theory of relativity over a range of phenomena
from the classical to the quantum, over distances from millimeters to the
astronomical, on time scales from the present to the early universe and
from the weak to the strong-field approximation.
An overall review of experimental gravitation was provided by Braginsky,
progress reports on resonant and laser gravitational wave detectors were
given by Bassan and Kawashima, respectively.  The status of tests of
the universality of free fall was presented by Adelberger and Boynton, some
new results as well as a review of previous experiments on gravitational
effect on superconductors and the quantum mechanical phase shift due
to gravity was
given by
Anandan and Werner. Paik gave a comprehensive description of superconducting
accelerometers and recent results from a null test of the inverse square
law of gravity. Results from binary pulsars, cosmic background radiation
and gravitational lensing was reported by Wolszczan, Partridge and Swings.
There were several interesting contributed papers also.
Details of the contents of this report can be found in the Proceedings
referred to earlier.

\vfil
\eject
\centerline{\bf  10th  Pacific Coast Gravity Meeting}
\medskip
\centerline{Jim Isenberg, Oregon}
\centerline{jim@newton.uoregon.edu}
\bigskip
\bigskip

The pre-eminent centers for General Relativity on the West Coast of
the US are at Caltech and at UC Santa Barbara. About ten years ago,
some of us from out near scri decided that it would be nice to hear
about the research that the people at those two places were doing;
and it would especially nice if this could be done on one trip.
Fortunately, the people at Caltech and UCSB also thought that this
was a good idea, and so with the support of Kip Thorne, Doug Eardley,
Jim Hartle, and Gary Horowitz, the first Pacific Coast Gravity
Meeting was organized and held at Caltech in March of 1985.

There was no time to get any grant money, so we decided that this
would be a near-zero budget conference, modeled after the "Steven's
Meetings" held years earlier on the East Coast. There were no invited
plenary speakers, and anyone who wanted to come could come; while
anyone who wanted to speak could speak. Talks by grad students were
especially encouraged. We also aimed to get people representing the
full breadth of GR and gravitational research--from astrophysics to
experimental measurements, and from numerical relativity to
mathematical relativity.

At the first PCGM, there were 33 talks going from Friday morning
until Saturday evening.  A very wide range of research was
discussed-- there were talks on quasilocal mass, on black hole
uniqueness theorems, on the newly popular superstring theory, on
rapidly rotating neutron stars, on gradiometer measurements, and on
competing schemes for detecting gravitational radiation. People
learned a lot, had a good time, and decided to do it again the next
year.

Pacific Coast Gravity meetings have been held (in February or March)
every year since then, with essentially the same format and even some
of the same faces, but with talks reflecting the changing interests
in GR. This year's PCGM was held in late March at Oregon State
University in Corvallis, Oregon. Tevian Dray and Corinne Manogue did
the bulk of the organization, and again we had an interesting and
enjoyable meeting, with about 80 participants listening to 36 talks
spread out from Friday morning until late Saturday afternoon.

One of the themes of this year's meeting was the need to prepare
theoretically for the coming of LIGO and its European counterpart
VIRGO. Just 10 days before the meeting, and just a couple of hundred
miles away from Corvallis, work had begun on site preparation for the
western branch of the laser interferometry gravity wave observatory,
in Hanford, Washington. Kip Thorne noted  that LIGO should be on the
air by 1998, and so he emphasized that if the theoreticians were
going to be ready to interpret the data it collects, they needed to
be calculating detailed profiles of the gravitational radiation which
one expects LIGO to see. Thorne indicated that  the primary sources
of the radiation which LIGO and VIRGO are expected to see are various
coalescing binary objects.  Millions of templates of these sources,
both for searching for radiation and for interpreting it, are needed,
and they must be calculated now. A number of researchers are already
working on this, and talks by Flanagan, Apostolatos, Laurance,
Poisson, and Kennefick discussed some of these results to date. The
talk by Price, on a new simple "hand calculator" study of colliding
black holes in the close limit, relates to this issue as well.

Another theme of this year's meeting was the success of the so-called
"EotWash"  experimental group in determining the accuracy of the
Equivalence Principle. Ever since the flurry of interest during the
mid 80's in the so called "5th Force", experimentalists have been
very carefully looking for any deviations from the Equivalence
Principle which might occur. Various distance ranges have been
explored, and various disparate materials have been examined.
Adelberger's  "EotWash" group in Seattle has done much of this
careful work, and he together with his students Smith, Gundlach, and
Harris reported on their results. So far, the Equivalence Principle
appears to be accurate to all orders, in all circumstances.

Quantum gravity has been a part of every one of the ten PCGM's. At
this year's meeting, the emphasis was on semiclassical studies,
especially near black holes. Hiscock talked about extreme black
holes, and how their special properties could be used to produce a
possible solar-sized laboratory for quantum gravity studies.
Hiscock's students--Bruekner, Whitesell, Herman, and
Loranz--discussed semiclassical analyses in Reissner-Nordstom
interiors, in black holes with scalar fields, and in 2-dimensional
black holes.  Other talks related to quantum gravity included the one
by Brooks which discussed the problem of finding observables, with
topological field theories used as a laboratory for studying this
problem; and the one by Cosgrove on 2+1 quantum gravity and the
possibility of topology change.

Many of the astrophysics talks at the Corvallis meeting focussed on
the problem of radiation production by coalescing binaries, as noted
above. There were other astrophysics-related talks as well. Ipser
discussed low frequency modes of rapidly rotating (periods of about 1
to 10 seconds)  systems, and showed that neutron stars with such
periods are consistent.  Nordtvedt described some celestial mechanics
studies of the lunar orbit, noting that in principle the lunar orbit
could be one of the best laboratories for studying nonlinear GR
effects. Das presented some exact solutions which describe spherical
collapse of an anisotropic body into a black hole.  A more formal
mathematical treatment of spherical GR systems was discussed by
Romano.

Topics ebb and flow. At the PCGM  four years ago in Eugene, there was
a lot of discussion of whether an arbitrarily advanced civilization
could build a time machine. At last year's meeting in Santa Barbara
there were no talks on this topic, but at this year's meeting, there
were two: Isenberg discussed a result which disproves the "Fountain
Conjecture"  of Hawking and Thorne (The Fountain Conjecture plays a
role in Hawking's argument that time machines can't be built). And
Tanaka did a semiclassical study of scalar fields which he says
indicates that quantum divergence may not work as a device to prevent
the construction of these machines. Superstring theory, too, has
provided lots of talks at some PCGM's and not so many at others. This
year, there was just one superstring talk: that by Horne, on
S-duality and chaos.  Cosmology, which has been well-represented at
many meetings, was really only discussed in one talk at Corvallis:
Agnew discussed certain exact cosmological solutions of Einstein's
equations with a stiff (nonperfect) fluid, and with planar symmetry.

One mainstay of all the PCGM's is Joe Weber. He reminded us that the
cylindrical resonant bar detectors at Maryland, Rome, and elsewhere
have been on the air for years, searching for gravitational
radiation.

A number of the talks at the meeting this year involved spinors and
Clifford algebras. Both Dray and Pezzaglia used Clifford algebra
ideas to get double sets of spinors, and each described how they
might be used in physics. Differ proposed a scheme for representing
all physical fields in terms of certain Clifford algebra-valued
fields. Taub described some features of spinors related to O(3,C) as
compared to those related to SL(2,C). And Schray discussed some
possible uses of octonionic spinors.

There were three other talks at the 10th PCGM. Boersma discussed some
of the mathematics which one encounters in studying spacetimes as
evolving local spatial geometries seen by a congruence of observers.
Pinto described how one might experimentally probe time-dependent
gravitational fields by examining atomic transitions in Rydberg
atoms near massive stars stars or neutron stars.
And Isaacson gave an Economics 101 seminar on the changing economic
realities at the NSF.  There is money at the NSF. But it may go
increasingly to areas of research which are deemed "strategic" to the
future of the nation. (Congress gets to decide what is "strategic"
and what isn't.)  Isaacson noted the need for us to let Congress know
why we think the sort of gravitaional research we do is useful and
deserving of support.

This coming year, on its 10th anniversary, the Pacific Coast Gravity
Meeting will return to Caltech, its place of birth. We warmly invite
you to come; if not this year, then some February or March in the
future.

\bye